\documentclass[preprint,12pt
]{elsarticle}

\usepackage[numbers]{natbib}
\def\tsc#1{\csdef{#1}{\textsc{\lowercase{#1}}\xspace}}
\tsc{WGM}
\tsc{QE}

\usepackage[toc,acronym]{glossaries}
\newacronym{AI}{AI}{Artificial Intelligence}
\newacronym{AIS}{AIS}{Axon Initial Segment}
\newacronym{AIMC}{AIMC}{Analog In-Memory Computing}
\newacronym{ANN}{ANN}{Artificial Neural Network}
\newacronym{ALU}{ALU}{Arithmetical and Logical Unit}
\newacronym{AP}{AP}{Action Potential}
\newacronym{APTD}{APTD}{Action Potential Time Derivative}
\newacronym{GPU}{GPU}{Graphic Processing Unit}
\newacronym{FPGA}{FPGA}{Field Programmable Gate Array}
\newacronym{CNN}{CNN}{Computer Neural Network}
\newacronym{CPU}{CPU}{Central Processing Unit}
\newacronym{HH}{HH}{Hodgkin and Huxley}
\newacronym{HW}{HW}{hardware}
\newacronym{HPCG}{HPCG}{High Performance Conjugate Gradients}
\newacronym{HPL}{HPL}{High-Performance Linpack}
\newacronym{ISI}{ISI}{Inter-Spike Interval}
\newacronym{I/O}{I/O}{Input/Output}
\newacronym{HPC}{HPC}{High Performance Computing}
\newacronym{LLM}{LLM}{Large Language Model}
\newacronym{ML}{ML}{Machine Learning}
\newacronym{OS}{OS}{operating system}
\newacronym{PD}{PD}{Propagation Delay}
\newacronym{OPS}{OPS}{Operations Per Second}
\newacronym{SOPS}{SOPS}{Synaptic Operations Per Second}
\newacronym{PSP}{PSP}{Post-Synaptic Potential}
\newacronym{PU}{PU}{Processing Unit}
\newacronym{SPA}{SPA}{Single Processor Approach}
\newacronym{SW}{SW}{software}
\newacronym{SNN}{SNN}{Spiking Neural Networks}

\begin{document}
\let\WriteBookmarks\relax
\def\floatpagepagefraction{1}
\def\textpagefraction{.001}



\title 
{	Do ions have a coating\\in neuronal electrolytes under an electric field?
	}  



%

\author[1]{János Végh}



\ead{Vegh.Janos@gmail.com}



\affiliation[1]{organization={Kalimános Bt
		},
            city={Budapest},
            postcode={1021}, 
            country={Hungary}}

\cortext[1]{Corresponding author}



\begin{abstract}	
Ions in electrolytes can have coatings and can combine into different complexes that significantly affects their size, mass, and transport features. Measuring such coatings of ions traveling in narrow,  limited spaces inside biological objects is challenging.
We assumed that the coating changes the size of the ion and that the speed of the complex may influence its composition.
The original Stokes-Einstein relation for diffusion assumes a simple, spherical particle in a homogeneous Newtonian fluid.
It was modified to describe an electric-field-driven drift. One possible application of the result is describing the propagation of axonal impulses, especially since soliton theory models it as a mechanical vibration, i.e., its speed changes rapidly. Hodgkin and Huxley measured the membrane current (due to axonal current) as a function of the clamping voltage, and they confirmed the linear dependence, that is, the independence of the ions' coating state from their speed.
\end{abstract}

%
%
%


\maketitle

\pagebreak
\section{Introduction}\label{sec:Introduction}

An electric field exerts a force on charged particles, leading to an additional drift velocity component in addition to the random thermal motion described by the Stokes-Einstein relation, which can introduce an electric field-driven drift. The motion of particles in an electric field is often described by a generalized diffusion equation that accounts for both random Brownian motion (thermal diffusion) and the directed drift caused by the electric field (electro-osmosis). The original Stokes-Einstein relation assumes a simple, spherical particle in a homogeneous Newtonian fluid. An electric field in fluids inside neurons can create complex interactions that can violate these assumptions and lead to deviations from the standard relation. 

\section{Membrane current\label{sec:experimental_axon_current}}

Below, I omit the current through the ion channels in the membrane's wall and the currents through other axons.

\subsection{Charge transfer vs current\label{subsec:Charge-transfer-vs-current}}

When describing the macroscopic phenomenon "current" in microscopic
terms, we apply a potential difference
to a macroscopic piece of space (or measure it) and measure the statistical
time course of the flow of the charge carriers. These two characteristics are
connected through the features of the medium (material) that hosts
our measurement. In a conducting wire, there are free charges; their
number per unit volume is given by $n$, and $q$ is the amount of
charge on each carrier. If the conductor has a cross-section of $A$,
in the length $dx$ of the wire, it has charge $dQ=q*n*A*dx$. If
the charges are moving with a macroscopic speed $v=\frac{dx}{dt}$, we define
the current as the charge moved per unit time:
\begin{equation}
	I=\frac{dQ}{dt}=q*n*A*v\label{eq:DriftCurrent}
\end{equation}
Notice that the macroscopic current $I$ is zero if any of the factors is zero. \emph{Microscopic carriers must be present in the volume}
and have charge, the cross-section must not be zero, and the charge
carriers must move with a velocity, which usually needs an external
force field. According to Stokes' Law, to move a spherical object
with a radius $R$ in a fluid having dynamic viscosity $\eta$, we need
a force
\begin{equation}
	F_{d}=6*\pi*\eta*R*v
\end{equation}
\noindent (drag force) acting on it (the effect of electric repulsion must be accounted for. The repulsion increases the driving force or virtually decreases viscosity). A (microscopic) electric force
field $\frac{dV}{dx}$ inside the wire would accelerate the charge
carriers continuously:
\begin{equation}
	F_{e}=k\frac{dV}{dx}q
\end{equation}
However, the medium in which the charge moves shows a (macroscopic,
speed-dependent) counterforce $F_{d}$, which in steady state equals
$F_{e}$, that is :
\begin{equation}
	I=\frac{k*q^{2}*A}{6*\pi*\eta*R}*\mathbf{n*\frac{dV}{dx}}\label{eq:StokesCurrent}
\end{equation}
The current in a wire is influenced by the electric force field (specific resistance) and the number of charge carriers $n$. While the latter is commonly considered constant and
part of the former, this is not necessarily true for biological
systems with electrically active structures inside. The medium's internal
structure introduces significant modifications. Applying an electric
field to a wire can generate varying amounts of current as the number
of charge carriers changes. For axons, we use a single-degree-of-freedom
system, a viscous damping model, so that the \emph{ions will move with
	a field-dependent constant velocity in the electric space}. However,
the activity of potential-controlled ion channels in the medium's wall may
change $n$ in various ways; furthermore, that change can result in
'delayed' currents during measurement, for example, in clamping: the $n$ gradually increases as ions diffuse into the tube.

\subsection{Axonal current\label{subsec:Axonalcurrent}}

We model the axons as electrolyte-filled semipermeable membrane tubes with ion channels in their walls.
Clamping
sets up an artificial working regime for the ion channels: the switched-on permanent
electric field on the outer surface opens the ion channels and enables ions to enter the inner volume where
formerly no ions (and, consequently, no potential) existed. The entering ions will
flow away from the place of their entrance in the direction of the gradient (that is, the current removes part of the ion layer on the surface), and a slow current toward the membrane can start. 
Under clamping conditions, the experimenter sets the voltage rather than the transmitted signal, and in a static rather than an autonomous dynamic way.

Initially, the membrane, the clamping point on the
axon, and the intracellular and extracellular fluid have the same
resting potential. When an external potential 
is applied  suddenly to some point of the axon, an electric field $\frac{dV}{dx}\propto(V_{membrane}-V_{clamp})$
appears on the \emph{outside surface} of the axon. The extracellular
space with its high ion concentration $C_{k}^{ext}$ represents an "ion
cloud". When the clamping voltage is switched on, a ``fast'' current
instantly delivers the potential along the \emph{outer} surface of
the axon. However, this is not the case (at least not in the initial
moment) on the \emph{inner} surface. Inside the axon, there is no charge present that could
change the potential: 'the intracellular concentration at rest is around
five orders of magnitude less than that in the extracellular space'~\cite{KochBiophysics:1999}. The physical picture that the clamping potential
instantly appears at the end of the axon at the membrane, i.e., if (apparently) they
have an infinitely large propagation speed, is wrong.

\begin{figure}
	\centering
	\includegraphics{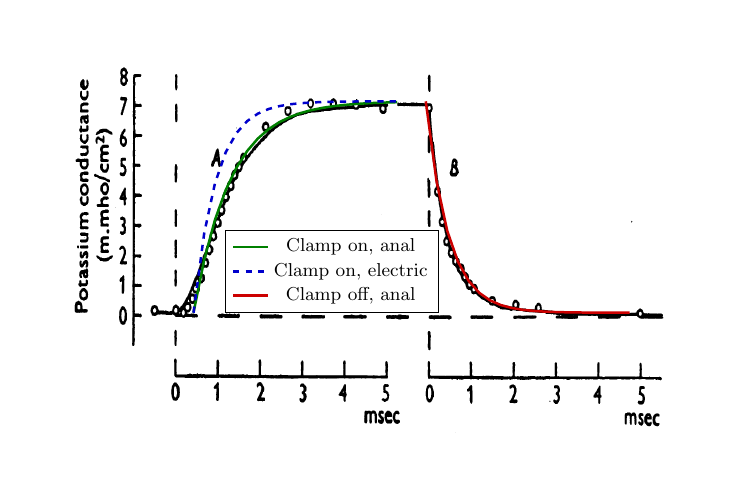}
	\caption{Measurement results from~\cite{HodgkinHuxley:1952}~Fig.~2, showing the asymmetrical 'charge and discharge' processes\label{fig:HHFig2Simple}}
\end{figure}

The persisting clamping voltage triggers the opening of ion channels in its wall along the axon,
leading to a continuous inflow through the axon's
wall from the extracellular space into the intracellular space as a "fast current".
The ions entering the intracellular space remain inside the axon: the cylindrical surface allows only one-way (inward) ion flow. "Once
calcium enters the intracellular cytoplasm it is not free to diffuse"~\cite{KochBiophysics:1999}. The ions start to create an ion-rich layer on the internal surface, that is, they create a
gradient parallel to the wall. The ions experience the electric field (which is present initially
only at the clamping point but extends with the passing time) along the axis, speed up, and (after a short while) the ion's
speed becomes constant in time. However, its value depends on the actual electric
field, see Equ.~(\ref{eq:DriftCurrent}). The \emph{ions will slowly
	move }along the axon \emph{with a field-dependent constant velocity
	in the electric space} in a viscous solution.
The moving ions deliver charge, so
the potential gradually extends along the electrolyte tube (the axon).
``In axon fibers, the
effective diffusion constant was estimated to be about one-tenth of
the diffusion coefficient in aqueous solution''~\cite{KochBiophysics:1999}; however, under the effect of the potential gradient, and the mutual repulsion,
they form a ``slow current'' (and that macroscopic current may have a much higher propagation speed). 
\emph{The current and the potential are not instant, as we consider in
	the classic theory of electricity}: they propagate with the speed
of the charge carriers.

In this model, we assume that during the time $dt$, in the volume
$dx$, we have a constant ion inflow $I_{wall}$ through the axon's
wall, which increases the charge and concentration already in the
volume. The charges in the tube experience the field $\frac{dV}{dx}$,
and they move with speed $v$ inside the tube (see Eq.~(\ref{eq:StokesCurrent})).
The ionic fluid with velocity $v$ (proportional
to $\frac{dV}{dx}$) transfers the ionic charge in
the volume to the neighboring element at a distance $v*dt$, and delivers
the charge and concentration from the neighboring element at a distance $-v*dt$ into
this element. At the time $t$, the concentration at $x$ will result
from the inflow at the place $x-v*t$. 
The higher the speed 
$v$, the more significant the difference between the "inflow" and the "present"
concentration. 
\begin{equation}
	\frac{dI_{axon}}{dt}=-\alpha*I_{axon};\ I_{axon}(t)\approx I_{wall}*(1-\exp(-\alpha*t))\label{eq:AxonalCurrentNoConc}
\end{equation}

\noindent ($\alpha$ is a timing constant
of dimension $(1/time)$).

\begin{figure}
	\centering
	\includegraphics{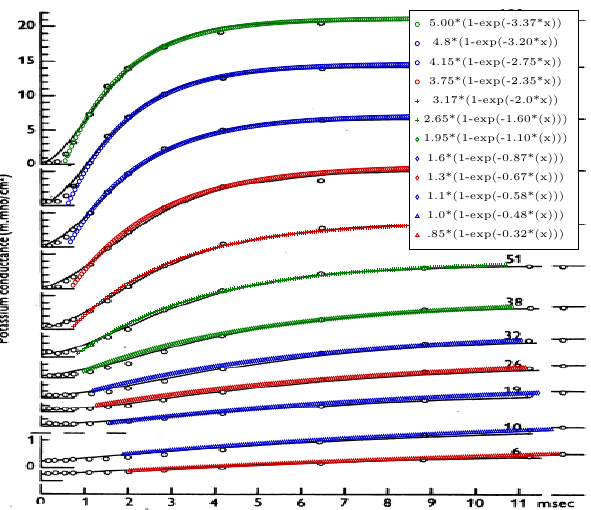}
	\caption{Measurement results from~\cite{HodgkinHuxley:1952}~Fig.~3, with fitting polynomial and exponential functions\label{fig:HHFig3Fitted}}
\end{figure}

The famous and influential axon current measurement
has been published in 1952~\cite{HodgkinHuxley:1952}. They used a single-axon input and measured the
neuronal membrane current, which was identical to the axon current. The diffused-in ions were transported towards
the membrane as a ``slow'' macroscopic ionic current (the speed
of current \gls{HH}~\cite{HodgkinHuxley:1952} measured
and also theoretically derived to be about $20~m/s$; it was in the
order of magnitude of the speed of macroscopic currents
in metals and electrolytes).

Fig.~\ref{fig:HHFig2Simple} shows the experimental evidence 
of the process. \gls{HH} measured the time course of the neuron membrane's current by
switching on and off an axonal current that charges the membrane. \gls{HH} modelled the membrane as a capacitor.
They noticed that the time constants of the charging and discharging processes are different, but they could not explain why. Actually, they used a charging current described by Eq.~(\ref{eq:AxonalCurrentNoConc}) and measured the resultant profile instead of charging the condenser with a constant current. This way, another exponential term comes into play, resulting in a markedly different charge-up time constant.

\begin{figure}
	\centering
	\includegraphics[scale=1.2,trim= 0 1.2cm 0 0,clip]{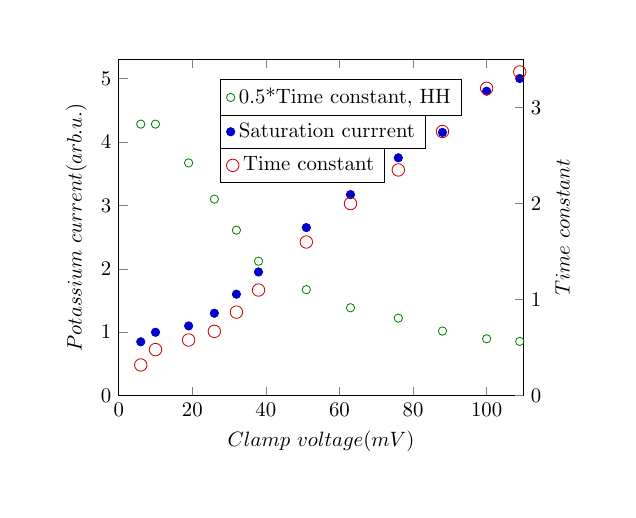}
	\caption{Evaluating data in Fig.~\ref{fig:HHFig3Fitted} with the 
		original polynomial and our exponential function}\label{fig:HHFig3linearity}
\end{figure}

They measured the time course of the axonal current (although they, following their predecessors~\cite{COLE_CURTIS_IMPEDANCE:1939} by mistake called it 'impedance'), at different clamping voltages. In addition to providing experimental evidence for the existence of "slow current", they also found that its speed dependence follows the Stokes-Einstein relation. Their result is reproduced in Fig.~\ref{fig:HHFig3Fitted}. 
As we discussed, ions diffuse in the axon's wall, producing a saturation-type current.
\gls{HH} fitted the experimental data with a polynomial, we use the theoretical function Eq.~(\ref{eq:AxonalCurrentNoConc}).
Our simple
model assumes that $\alpha$ is constant in time but (through $v$
and $\frac{dV}{dx}$) depends on the clamping voltage.
We expect (see Eq.(\ref{eq:StokesCurrent})) that the value of the time constant and the value of the saturation current depend linearly on the clamping voltage.  Fig.~\ref{fig:HHFig3linearity} shows the
saturation current and the time constant derived by fitting those functions to the experimentally measured data.

Charge carriers flowed in through the walls continuously, the ``slow''
current ``flowing out'' from the tube increased and appeared on the
membrane, \emph{establishing the illusion that the conductance of
	the tube increases}. \emph{Resistance/conductance cannot be interpreted
	when charge carriers flow into the resistor} (and this is the case
with the axon) from the environment: the ``slow'' current produced
a different behavior (increased the number of charge carriers $n$), see Eq.~(\ref{eq:StokesCurrent}).

\gls{HH}~\cite{HodgkinHuxley:1952}
were thinking in the Newtonian way; that is, they assumed that the transport speed of electrodiffusion in an axon and the electromagnetic field propagation were the same,  so \emph{the slowly moving
	charged object that they observed could not be a current}. They expected
the current to appear promptly after switching on the clamping voltage.

As Fig.~\ref{fig:HHFig3Fitted} displays, a systematic discrepancy exists at the low time values
between the time courses of the measured data and the polynomial function fitted originally by~\cite{HodgkinHuxley:1952}, while 
our saturation curve provides a much better fit.
The former one is a wrong quasi-model; our fitting uses the correct model function. The dependence we use (a sudden exponential increase in the membrane's current) has been experimentally 
measured by~\cite{SodiumCurrentDelay:2006}.
When using our data evaluation method, the figure suggests that the saturation current depends on the ions' speed (i.e., the clamping voltage; see Fig.~\ref{fig:HHFig3Fitted} and Fig.~\ref{fig:HHFig3linearity}) in the tube.

\subsection{Evaluating the experimental data\label{subsec:EvaluatingExperiment}}

\gls{HH} have measured that the time constants in the
exponentials are $1.1\ ms$ and $0.75\ ms$, respectively, 
for their Fig.~2. Those time constants are markedly different: the
first one describes the ``slow'' current (decreased by macroscopic
ion flow), the second one the net ``fast'' current. 
This disagreement alone should have called attention to the fact that the time constants belong to different physical processes.

The figure also calls attention to the importance of 
using the correct data evaluation method and measurement method.
\gls{HH} used a polynomial to fit their data. (Huxley used a mechanical calculator~\cite{CompanionGuideHodgkinHuxley:2022}, so he wanted to avoid calculations of functions such as an exponential. That oversimplification should not be relevant in the age of modern electronic computers).
A polynomial is fitted to the constant current values at low time values, plus the correct exponential function~\cite{SodiumCurrentDelay:2006}, which distorts
the theoretical values at those time values.  
The lower speed in the case of low clamping voltage 
resulted in longer 'no current' region due to the use of the electrolyte 
electrodes, as shown in Fig.~\ref{fig:HHFig3Fitted}.
This effect 
is so significant that the values of the time constants derived by \gls{HH} are not even
close to the correct ones derived by the
correct theoretical function. Furthermore, their tendency suggests
a dependence opposite to the correct one, see Fig.~\ref{fig:HHFig3linearity}.
Since the linearity persists at low speeds (although dedicated measurements are needed), one can hypothesize that the ions
do not wear coat in neuronal electrolytes.

Another aspect is to measure the correct physical parameter.
Fig.~\ref{fig:HHFig3Fitted} is told to show the conductance,
but actually it measures the sum of the current from the 
measuring device plus the ion current.
The principle of measuring conductance is, see \cite{JohnstonWuNeurophysiology:1995}, section
A.3.12: "\textit{(input impedance) can be measured by applying a voltage
	and measuring the resulting current or by \textbf{injecting a current} and
	measuring the resulting voltage}".
The relative contribution of the measuring current increases as the ion current decreases. Its effect 
results in that in Fig.~\ref{fig:HHFig3linearity} the saturation current is well above the expected
linearity at low ion current values, while it shows 
linear behavior at higher current values.

\section{Summary}\label{sec:Summary}

The famous experiment by \gls{HH} also derived experimental evidence
that the ions in the electrolytes of the neuron obey the Stokes-Einstein relation. Their experimental data underpin that the ions' water-related 
structure is independent of their speed; furthermore, that 
the electric field due to clamping, instead of some ion channel mechanism, moves the ions.
To discover that relation,
one must apply the correct theoretical model, fix the 
data evaluation and measurement data interpretation errors.
Without doing so, the inappropriate data evaluation method derives entirely wrong conclusions from the correct measured data.


%



\end{document}